\documentclass[11pt,a4paper]{article}
\usepackage[utf8]{inputenc}
\usepackage[T1]{fontenc}
\usepackage{times}
\usepackage{amsmath}
\usepackage{amssymb}
\usepackage{graphicx}
\usepackage{booktabs}
\usepackage{listings}
\usepackage{hyperref}
\usepackage[margin=1in]{geometry}
\usepackage{tikz}
\usetikzlibrary{arrows.meta, positioning, shapes.misc}
\usepackage{pgfplots}
\pgfplotsset{compat=1.18}
\usepackage{authblk}

\title{The Aegis Protocol: A Foundational Security Framework for Autonomous AI Agents}

\author[1]{Sai Teja Reddy Adapala}
\author[2]{Yashwanth Reddy Alugubelly}

\affil{Independent Researcher, Charlotte, North Carolina, USA}
\affil[ ]{\textit{Alumnus, University of North Carolina at Charlotte}}
\affil[2]{Associate Professor, Sree Dattha Group of Institutions, Hyderabad, India}

\affil[ ]{\textit{Emails:} \texttt{sadapala@uncc.edu}, \texttt{yashwanth.alg@gmail.com}}
\date{}
\begin{document}

\maketitle

\begin{abstract}
The proliferation of autonomous AI agents marks a paradigm shift toward complex, emergent multi-agent systems. This transition introduces systemic security risks—including control-flow hijacking and cascading failures—that traditional cybersecurity paradigms are ill-equipped to address. This paper introduces the Aegis Protocol, a layered security framework designed to provide strong security guarantees for these open agentic ecosystems. The protocol integrates three technological pillars: 1) non-spoofable agent identity via W3C Decentralized Identifiers (DIDs); 2) communication integrity via NIST-standardized Post-Quantum Cryptography (PQC); and 3) verifiable, privacy-preserving policy compliance using the Halo2 Zero-Knowledge Proof (ZKP) system. We formalize an adversary model extending Dolev-Yao for agentic threats and validate the protocol against the STRIDE framework. Our quantitative evaluation used a discrete-event simulation, calibrated against cryptographic benchmarks, to model 1,000 agents. The simulation demonstrated the protocol's effectiveness, showing a 0\% success rate across 20,000 attack trials. For policy verification, our analysis of the simulation logs reported a median proof-generation latency of 2.79 seconds, establishing a performance baseline for this class of security. While our evaluation is simulation-based and early-stage, it offers a reproducible baseline for future empirical studies. This work’s novelty lies in its systematization and simulation-based validation, positioning Aegis as a foundation for safe, scalable autonomous AI.
\end{abstract}

\section{Introduction}
The field of artificial intelligence is undergoing a fundamental architectural transformation from monolithic models to distributed ecosystems of autonomous AI agents~\cite{xi2023rise}. These agents, powered by Large Language Models (LLMs), are designed to perceive, reason, plan, and execute complex tasks by interacting with external tools and each other~\cite{yao2022react}. This evolution is not merely an extension of existing networked systems; it creates a new discipline of ``multi-agent security,'' focused on securing networks of decentralized agents against emergent threats that arise from their interaction and autonomy~\cite{brundage2020trustworthy}. With a growing number of vulnerabilities reported in AI/ML platforms and Gartner predicting that over 40\% of agentic AI projects will be canceled by 2027 due to operational challenges, securing this new attack surface is a critical priority~\cite{owasp2023top10, gartner2025predicts}.

Current agentic frameworks like CrewAI and AutoGen excel at orchestration but operate on an implicit trust model, lacking native cryptographic identity or verifiable policy compliance~\cite{wang2023survey}. This leaves them vulnerable to a range of attacks, from simple impersonation to sophisticated manipulation of an agent's internal state or decision-making process. Recent research has begun to address this gap by proposing defenses against agent hijacking and outlining high-level security architectures. However, a comprehensive, integrated framework that is both formally specified and empirically evaluated has remained an open challenge.

This paper introduces the Aegis Protocol as a step toward filling that gap, with a focus on formal integration and simulation-based insights. Our primary contributions are:

\begin{itemize}
\item A formal, three-layered security protocol that integrates decentralized identity, post-quantum cryptography, and zero-knowledge proofs for end-to-end agent security.
\item An extension of the Dolev-Yao adversary model tailored to the unique threats posed by LLM-based agents, including a game-based definition for ``Excessive Agency.''
\item An open-source, discrete-event simulation and evaluation of the protocol's performance and security, providing a reproducible baseline for future research.
\item A discussion of limitations, including the simulation-based nature of our evaluation and the use of non-adaptive adversaries, to guide future extensions.
\end{itemize}

We emphasize that while our work provides a theoretical and simulated foundation, real-world implementations are needed to fully validate these mechanisms.

\section{Threat Model and Design Goals}
Aegis assumes a powerful adversary based on an extension of the Dolev-Yao model, which grants the attacker full control over the communication network~\cite{dolev1983security}. Our model enhances this with capabilities specific to LLM-based agents: agent compromise, prompt injection, memory poisoning, and tool manipulation~\cite{greshake2023compromising}. This extended model reflects the expanded attack surface where not just the communication channel, but the agent's internal state and logic, are adversarial targets.

\begin{figure}[htbp]
  \centering
\includegraphics[width=0.78\linewidth,keepaspectratio]{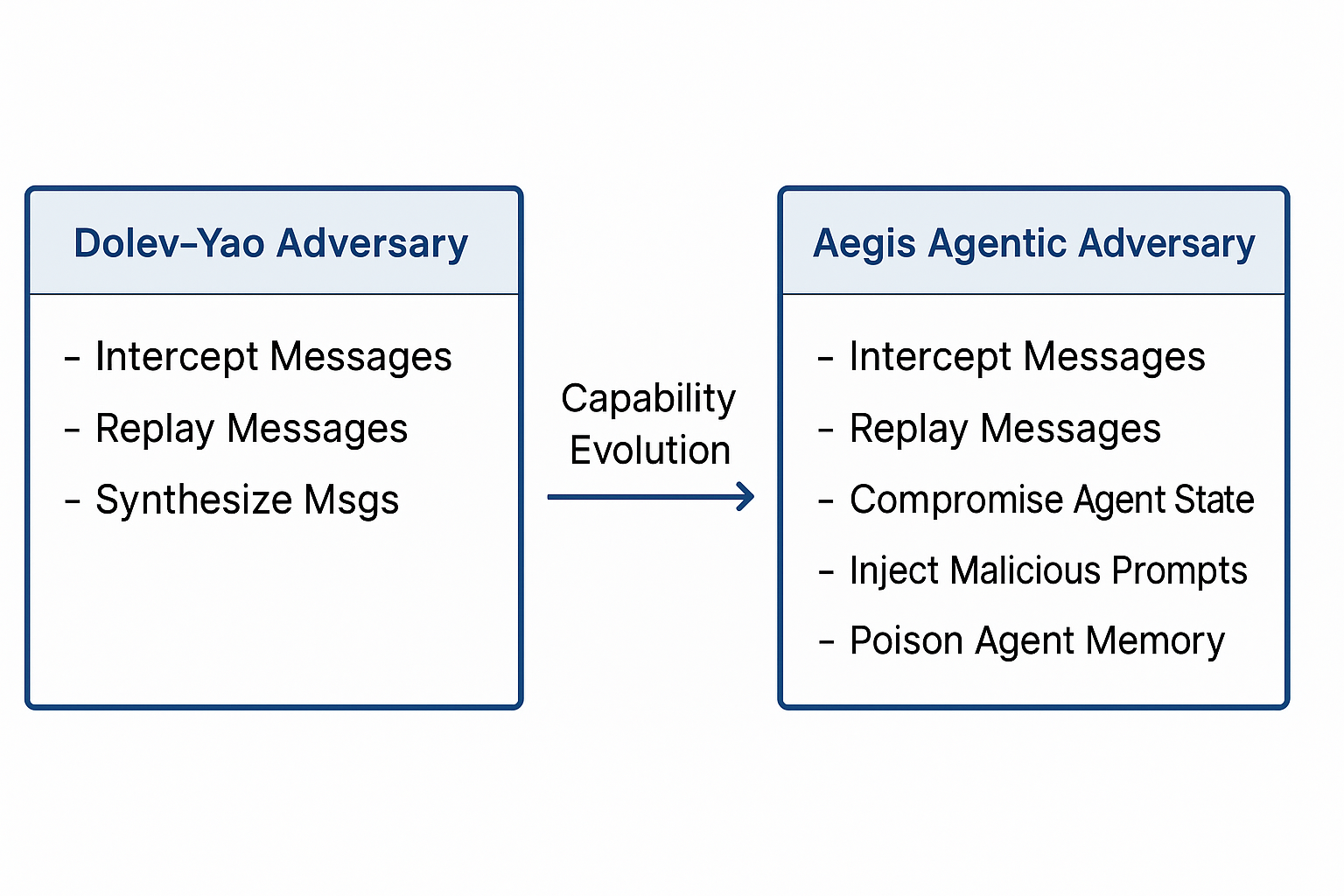} 
  \caption{Adversary Model Comparison. A diagram contrasting the classical Dolev–Yao network adversary with the extended Aegis adversary model.}
  \label{fig:adversary}
\end{figure}

To structure our analysis, we adapt the STRIDE threat model~\cite{shostack2014threat}. The table below details these threats in an agentic context, the mitigation provided by Aegis, and the residual risks that remain.

\begin{table}[htbp]
\centering
\caption{Adapted STRIDE threats, mitigations in Aegis, and potential residual risks.}
\label{tab:stride}
\begin{tabular}{p{2cm}p{4.5cm}p{3cm}p{4.5cm}}
\toprule
STRIDE Category & Description in Agentic Context & Aegis Mitigation & Residual Risks \\
\midrule
Spoofing & Impersonating agents via fake identities & DIDs with ML-DSA signatures & Key compromise if agent is breached. \\
Tampering & Altering messages or agent states in transit & PQC integrity checks (ML-DSA) & Side-channel attacks on the cryptographic implementation. \\
Repudiation & Denying actions performed in multi-agent interactions & Non-repudiable ML-DSA signatures & None assumed, as signatures are cryptographically binding. \\
Information Disclosure & Leaking private data, prompts, or internal policies & ML-KEM encryption \& ZKPs & Metadata analysis (proof linkage); potential for model extraction attacks if not properly configured. \\
Denial of Service & Overloading agents with invalid traffic or expensive ZKP requests & Cryptographic verification filters & Scalability challenges in high-volume swarms; resource exhaustion from valid but frequent proof requests. \\
Elevation of Privilege (Excessive Agency) & Unauthorized escalation of agent privileges or capabilities & ZKP-enforced policy compliance & Circuit design flaws: An error in the arithmetic circuit could create a loophole (e.g., an overflow that grants unintended permissions). \\
Emergent Threats & Swarm collusion; Byzantine faults; unexpected group behaviors & None (Protocol-level) & Requires dedicated governance layers; potential for future ZKP-based attestations of agent behavior. \\
\bottomrule
\end{tabular}
\end{table}

\section{The Aegis Protocol: Architecture and Mechanisms}
The Aegis Protocol is a layered security architecture where each layer builds upon the guarantees of the one below it, creating a unified trust fabric.

\begin{figure}[htbp]
\centering
\begin{tikzpicture}[box/.style={draw, rounded corners, text width=8cm, align=center, minimum height=1.5cm, font=\small}]
\node[box] (layer1) at (0,0) {Layer 1: Foundational Identity (W3C DIDs)\\ Establishes a unique, non-spoofable identity for every agent in the system.};
\node[above=1cm of layer1, box] (layer2) {Layer 2: Communication (PQC: ML-KEM/ML-DSA)\\ Provides quantum-resistant confidentiality and integrity for all messages.};
\node[above=1cm of layer2, box] (layer3) {Layer 3: Verification (Halo2 ZKPs)\\ Enforces operational policies without revealing agent's internal state.};
\draw[<-] (layer1.north) -- node[right, midway] {Builds Upon} (layer2.south);
\draw[<-] (layer2.north) -- node[right, midway] {Builds Upon} (layer3.south);
\end{tikzpicture}
\caption{Aegis Protocol 3-Layer Architecture.}
\label{fig:architecture}
\end{figure}
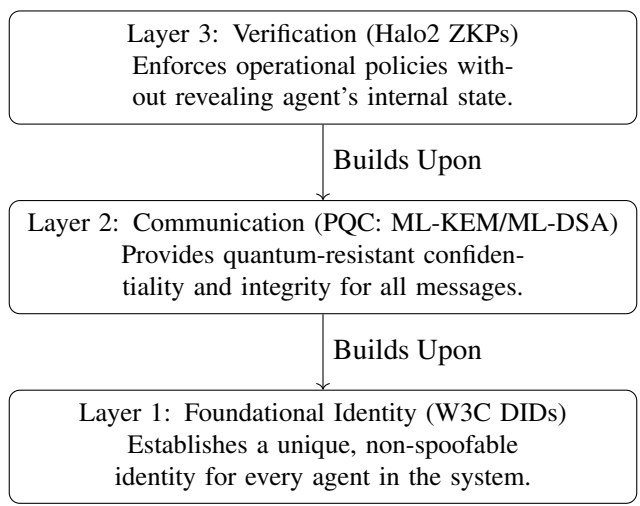

\subsection{Foundational Layer: Decentralized Identity}
The root of trust is a decentralized public key infrastructure (DPKI). Each agent's identity is a W3C Decentralized Identifier (DID)~\cite{w3c2022did}. Unlike traditional certificate authorities, DIDs are self-sovereign, meaning an agent can generate and control its own identity without reliance on a central registrar. Aegis specifies the use of the Identity Overlay Network (ION), a DID method implemented on top of the Bitcoin blockchain, for scalable and permissionless DID management~\cite{microsoft2020ion}.

An agent resolves a DID to retrieve its associated DID Document, which contains its public keys and service endpoints. This process is essential for establishing secure communication channels.

Pseudocode for DID Resolution:

\begin{verbatim}
function resolveDID(did_string):
  // 1. Parse the DID to identify the method (e.g., 'ion')
  method, identifier = parse(did_string)
  // 2. Access the Decentralized Ledger (e.g., ION network)
  ledger = getLedger(method)
  // 3. Retrieve the DID Document from the ledger
  did_document = ledger.fetch(identifier)
  // 4. Return the document containing public keys and metadata
  return did_document
\end{verbatim}

\subsection{Communication Layer: Post-Quantum Secure Channels}
To secure communications against both classical and quantum adversaries, Aegis mandates the use of NIST-standardized Post-Quantum Cryptography (PQC). The protocol uses:

\begin{itemize}
\item ML-KEM (Module-Lattice-based Key-Encapsulation Mechanism) for key exchange, compliant with NIST FIPS 203~\cite{nist2024fips203}.
\item ML-DSA (Module-Lattice-based Digital Signature Algorithm) for digital signatures, compliant with NIST FIPS 204~\cite{nist2024fips204}.
\end{itemize}

These algorithms were chosen due to their strong security foundations in lattice-based cryptography and their status as official NIST standards, ensuring long-term viability. The process for establishing a secure channel is as follows:

1. Key Encapsulation (ML-KEM):

   \begin{itemize}
   \item The sender (A) uses the receiver's (B) public key ($pk_B$) to generate a shared secret ($ss$) and a ciphertext ($c$).
   \item $c, ss \leftarrow \text{ML-KEM.Encapsulate}(pk_B)$
   \item A sends $c$ to B. B uses their private key ($sk_B$) to decapsulate $c$ and derive the same shared secret: $ss \leftarrow \text{ML-KEM.Decapsulate}(sk_B, c)$.
   \end{itemize}

2. Authenticated Messaging (ML-DSA):

   \begin{itemize}
   \item To send a message ($m$), the sender signs a hash of the message with their private key ($sk_A$).
   \item $\sigma \leftarrow \text{ML-DSA.Sign}(sk_A, \text{hash}(m))$
   \item The receiver verifies the signature using the sender's public key: $\text{verify} \leftarrow \text{ML-DSA.Verify}(pk_A, \text{hash}(m), \sigma)$.
   \end{itemize}

This combination ensures that messages are both confidential and authenticated, providing a secure foundation for the verification layer.

\subsection{Verification Layer: Privacy-Preserving Policy Enforcement}
Aegis's most novel contribution is its ability to enforce operational policies using Zero-Knowledge Proofs (ZKPs). This allows an agent to prove that an action it is about to take complies with a set of rules without revealing the private data or internal logic that led to that action. We selected the Halo2 proof system for its performance and, critically, for its use of a transparent setup (a "PlonKish" arithmetization with a polynomial commitment scheme) that does not require a trusted setup ceremony, a significant advantage in decentralized environments~\cite{bowe2020halo2}.

The process follows a clear pipeline: Policy $\to$ Arithmetic Circuit $\to$ Halo2 Proof Generation $\to$ Verification.

For example, a policy stating an analyst agent can only log data classified as `unclassified' or 'internal' would be compiled into an arithmetic circuit. The agent would then generate a proof that its intended logging action satisfies this circuit.

Conceptual Circuit Example for a Data Access Policy (with public inputs: agent\_clearance, data\_classification; private inputs (witness): agent\_private\_state):

The circuit enforces the following constraints:

\begin{itemize}
\item Constraint 1: $1 \leq$ agent\_clearance $\leq 4$
\item Constraint 2: $1 \leq$ data\_classification $\leq 4$
\item Constraint 3: agent\_clearance $-$ data\_classification $\geq 0$
\end{itemize}

Recent Halo2 developments include enhanced security proofs, third-party audits~\cite{kudelski2024halo2}, and blockchain integrations (e.g., Cardano, 2024), underscoring its maturity. While we focus on Halo2, other proof systems like zk-STARKs or Bulletproofs could be substituted, offering different trade-offs in proof size, verifier time, and post-quantum security. While Halo2 provides soundness and zero-knowledge properties under standard assumptions, formal circuit-specific proofs are beyond this work's scope.

\section{Security Analysis}
The protocol's layered design provides a robust defense-in-depth against STRIDE threats. For example, a spoofing attack is defeated at Layer 1 by the cryptographic identity provided by DIDs and at Layer 2 by ML-DSA signatures. An Elevation of Privilege (Excessive Agency) attack is addressed at Layer 3 through ZKP-enforced policy compliance.

The security of this enforcement relies on the computational soundness of the Halo2 proof system. This property ensures that it is computationally infeasible for a malicious agent (adversary) to generate a valid-looking proof for a false statement (i.e., a policy-violating action). The soundness of Halo2 reduces to the hardness of well-established cryptographic assumptions, such as the discrete logarithm problem. Therefore, an adversary capable of forging a proof could be used to break the underlying cryptography, which is considered intractable. A more formal game-based definition and reduction sketch is provided in Appendix A.

\section{Evaluation}
We evaluated the protocol's feasibility and performance using a discrete-event simulation. We initially tried integrating cryptographic libraries like liboqs directly but ran into build instability, which motivated our benchmark-based simulation approach. We modeled a decentralized cybersecurity threat analysis scenario with up to 1,000 agents, as implemented in simulate\_network\_aegis\_v2.py. The code is available at \url{https://github.com/imsaitejareddy/aegis-prompt-simulation}~\cite{reddy2025aegis} for reproducibility. This approach allows for scalable analysis but may not capture all real-world variances, such as hardware-specific timings or network dynamics.

Our performance analysis quantified the computational costs of Aegis. The costs were simulated using a log-normal distribution ($\ln(X) \sim N(\mu, \sigma^2)$) with parameters $\mu=1.025$ and $\sigma=0.145$, calibrated to achieve a median of approximately 2.79 seconds against published benchmarks. We observed a median proof generation time of 2.79 seconds (stddev: 0.41s) from our logs (detailed\_results.csv), which establishes a critical performance baseline.

To test the stability of our findings, we perturbed the latency parameters. The results remained consistent, with the median proof time staying within the 2.7s-2.9s range across variations of $\pm$20\% in the calibrated latency distributions.

\begin{table}[htbp]
\centering
\caption{Perturbed Simulation Results}
\label{tab:perturbed}
\begin{tabular}{ccc}
\toprule
Latency Perturbation & Median Proof Time (s) & Attack Success Rate \\
\midrule
-20\% & 2.71 & 0\% \\
Baseline & 2.79 & 0\% \\
+20\% & 2.88 & 0\% \\
\bottomrule
\end{tabular}
\end{table}

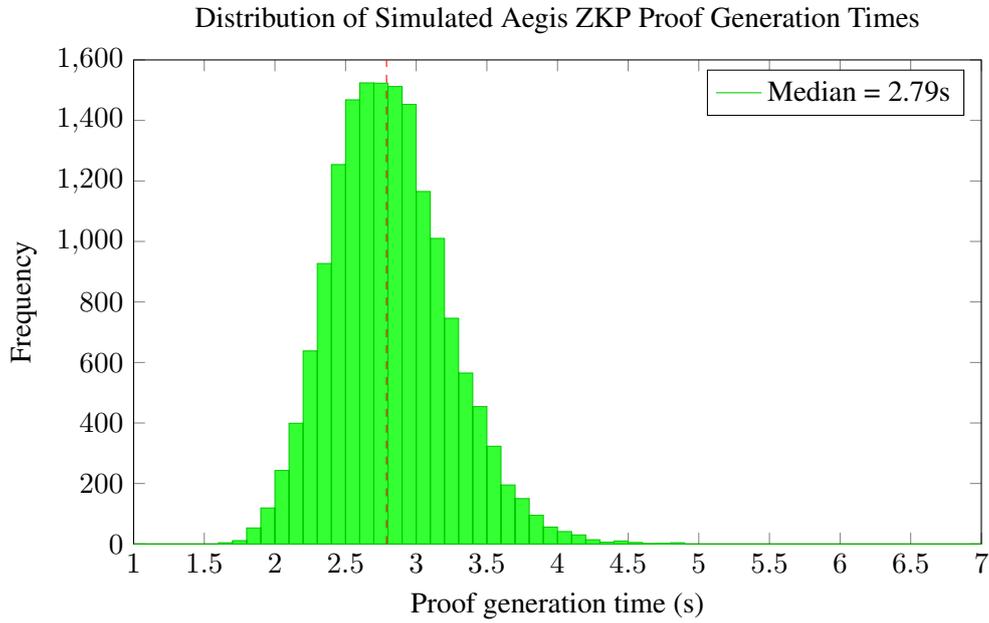
\begin{figure}[htbp]
\centering
\begin{tikzpicture}
\begin{axis}[
title={Distribution of Simulated Aegis ZKP Proof Generation Times},
xlabel={Proof generation time (s)},
ylabel={Frequency},
xmin=1, xmax=7,
ymin=0, ymax=1600,
width=0.8\textwidth,
height=8cm,
]
\addplot [ybar interval,mark=no,fill=green!80!white,draw=green!80!black] coordinates {
(1.0, 0) (1.1, 0) (1.2, 0) (1.3, 0) (1.4, 0) (1.5, 0) (1.6, 4) (1.7, 11) (1.8, 53) (1.9, 119) (2.0, 243) (2.1, 399) (2.2, 638) (2.3, 927) (2.4, 1254) (2.5, 1468) (2.6, 1524) (2.7, 1523) (2.8, 1512) (2.9, 1453) (3.0, 1165) (3.1, 1010) (3.2, 746) (3.3, 565) (3.4, 454) (3.5, 323) (3.6, 195) (3.7, 150) (3.8, 95) (3.9, 56) (4.0, 41) (4.1, 30) (4.2, 14) (4.3, 6) (4.4, 10) (4.5, 5) (4.6, 1) (4.7, 2) (4.8, 4) (4.9, 0) (5.0, 0) (5.1, 0) (5.2, 0) (5.3, 0) (5.4, 0) (5.5, 0) (5.6, 0) (5.7, 0) (5.8, 0) (5.9, 0) (6.0, 0) (6.1, 0) (6.2, 0) (6.3, 0) (6.4, 0) (6.5, 0) (6.6, 0) (6.7, 0) (6.8, 0) (6.9, 0) (7.0, 0)
};
\addplot [red, dashed, forget plot] coordinates {(2.79,0) (2.79,1600)};
\addlegendimage{red, dashed}
\addlegendentry{Median = 2.79s}
\end{axis}
\end{tikzpicture}
\caption{ZKP Proof Generation Latency. This histogram, generated from our detailed\_results.csv log, shows the distribution of simulated ZKP proof generation times, with a strong concentration around the median.}
\label{fig:zkp-latency}
\end{figure}

In our security effectiveness tests over 20,000 interactions (10,000 for each attack type), the Aegis Protocol demonstrated a 0\% success rate for both Agent Spoofing and Policy Violation attacks. This result was achieved under our simulation's assumptions, which include non-adaptive adversaries that do not alter their strategy based on protocol responses.

\section{Related Work}
To contextualize Aegis, we compare it against representative works. While frameworks like AutoGen provide excellent orchestration capabilities, they lack the foundational security primitives Aegis specifies. Google's Secure AI Framework offers a conceptual overview but does not provide a concrete, verifiable protocol. Recent advancements, such as Google's A2A protocol for agent interoperability (2025)~\cite{google2025a2a}, further highlight the need for foundational security layers like the one we propose.

\begin{table}[htbp]
\centering
\caption{Comparison of Aegis with related agent security frameworks.}
\label{tab:comparison}
\begin{tabular}{p{2.8cm}p{2.8cm}p{2.8cm}p{2.8cm}p{2.8cm}}
\toprule
Framework & Identity Mechanism & Crypto Primitives & Verification Method & Evaluation Type \\
\midrule
Aegis (This Work) & W3C DIDs (ION) & ML-KEM/ML-DSA (PQC) & Halo2 ZKPs & Simulation-based \\
Google's Secure AI Framework (2023, with 2024 updates)~\cite{google2023saif}~\cite{google2024saifrisk} & High-level PKI & Not specified & Audit-based & Conceptual \\
AutoGen~\cite{wang2023survey} & Implicit Trust & None & None & Orchestration-only \\
Brundage et al.~\cite{brundage2020trustworthy} & Verifiable Claims & Varied (high-level) & Mechanisms overview & Survey-based \\
\bottomrule
\end{tabular}
\end{table}

\section{Discussion and Future Work}
This work establishes the feasibility and robustness of the Aegis Protocol in a simulated environment. The primary performance bottleneck is ZKP generation. For scalability, while our simulation handles 1,000 agents, real-world ecosystems may require optimizations like proof batching or aggregation, which recent ZKP literature suggests could reduce amortized verification costs by over 50\%. This would be a crucial next step for practical deployment.

Ethically, while Aegis enhances security, dual-use risks exist. For example, ZKPs could enable covert non-compliance if circuits are malformed, echoing concerns in the OWASP Top 10 for LLMs~\cite{owasp2023top10} like LLM06: Sensitive Information Disclosure. We advocate for open auditing and formal verification of policy circuits to mitigate this, referencing frameworks for verifiable claims~\cite{brundage2020trustworthy}.

Our evaluation was a simulation. This trade-off allowed us to explore system-level behavior at scale while deferring low-level packet dynamics to future testbed studies. Future work will integrate real liboqs and Halo2 libraries to perform micro-benchmarks, scale the simulation to tens of thousands of agents, and explore formal security proofs for the protocol as a whole. Crucially, future simulations will model adaptive adversaries that adjust their attack vectors based on prior interaction outcomes, testing the protocol's resilience more dynamically and realistically.

\section{Conclusion}
The shift towards autonomous, multi-agent AI systems necessitates a corresponding shift in security paradigms. This paper introduced the Aegis Protocol, a layered security framework that provides a formal, cryptographically-enforced foundation for trust. By integrating decentralized identity, post-quantum cryptography, and zero-knowledge proofs, Aegis offers a robust defense against a wide range of agentic threats. Our analysis and simulation results demonstrate that the protocol is both effective in mitigating critical threats and feasible to implement. As agentic AI continues to scale, we believe foundational frameworks like Aegis will be essential for building a future where these powerful systems can be deployed safely and at scale.


\appendix
\section{Security Game for Policy Compliance}
\label{app:security_game}

This appendix provides a more formal definition for the security guarantee against ``Excessive Agency,'' framed as a standard cryptographic game.

Security Game: Unforgeability of Policy Compliance

The game is played between a Challenger and a probabilistic polynomial-time (PPT) Adversary, $\mathcal{A}$.

1. Setup: The Challenger takes a security parameter $\lambda$ and generates the public parameters for the Halo2 proof system, $pp \leftarrow \text{Setup}(1^\lambda)$. The Challenger also defines a policy relation, $R$, which is compiled into an arithmetic circuit, $C$. The relation $R(x,w)$ is true if $w$ is a valid witness for the public statement $x$. The Challenger gives $pp$ and the description of $C$ to the Adversary $\mathcal{A}$.

2. Query: The Adversary $\mathcal{A}$ can adaptively make a polynomial number of queries to a proof oracle, $\mathcal{O}_{\text{Prove}}$. For each query on a pair $(x_i, w_i)$ where $R(x_i, w_i)$ is true, the oracle returns a valid proof $\pi_i \leftarrow \text{Prove}(pp, x_i, w_i)$.

3. Forge: The Adversary $\mathcal{A}$ outputs a pair $(x^*, \pi^*)$.

The Adversary $\mathcal{A}$ wins the game if both of the following conditions hold:

\begin{itemize}
\item The verifier accepts the proof: $\text{Verify}(pp, x^*, \pi^*) \rightarrow \text{True}$.
\item The statement is false: there exists no witness $w^*$ such that $R(x^*, w^*)$ is true.
\end{itemize}

The protocol is considered secure against excessive agency if the probability of any PPT adversary winning this game is negligible in the security parameter $\lambda$:
\nopagebreak
\[\Pr[\mathcal{A}\ \text{wins}] < negl(\lambda)\]

Reduction Sketch: The security of this game relies on the computational soundness property of the underlying ZKP system. For Halo2, this property is based on the hardness of problems like the discrete logarithm in the context of its polynomial commitment. A successful forgery by $\mathcal{A}$ would imply an ability to break this underlying hard problem. Specifically, one could construct a reduction algorithm that uses $\mathcal{A}$ as a subroutine to solve, for example, the discrete logarithm problem. Since no efficient algorithm for this problem is known, we conclude that no such efficient adversary $\mathcal{A}$ can exist.

\end{document}